\documentclass[sigconf]{acmart}

\AtBeginDocument{%
  \providecommand\BibTeX{{%
    \normalfont B\kern-0.5em{\scshape i\kern-0.25em b}\kern-0.8em\TeX}}}

\copyrightyear{2022} 
\acmYear{2022} 
\setcopyright{rightsretained} 
\acmConference[SIGIR '22]{Proceedings of the 45th International ACM SIGIR Conference on Research and Development in Information Retrieval}{July 11--15, 2022}{Madrid, Spain}
\acmBooktitle{Proceedings of the 45th International ACM SIGIR Conference on Research and Development in Information Retrieval (SIGIR'22), July 11--15, 2022, Madrid, Spain}
\acmDOI{10.1145/3477495.3531997}
\acmISBN{978-1-4503-8732-3/22/07}

\usepackage[ruled]{algorithm2e}
\usepackage{amsmath}
\usepackage{color}
\usepackage{subfig}
\usepackage{balance}
\usepackage{ifthen}
\usepackage{multirow}
\usepackage{enumitem}
\usepackage[english]{babel} 

\usepackage[T1]{fontenc}
\usepackage[utf8]{inputenc}
\usepackage{babel}
\usepackage[font=small,labelfont=bf,tableposition=top]{caption}
\usepackage{booktabs}
\usepackage{threeparttable}

\definecolor{exampleblue}{RGB}{30,144,255}
\definecolor{examplepink}{RGB}{255,192,203}
\definecolor{exampleyellow}{RGB}{255,227,132}
\begin{document}
\fancyhead{}

\title{Incorporating Explicit Knowledge in Pre-trained Language Models for Passage Re-ranking}
\author{Qian Dong$^{1,3\dagger}$, Yiding Liu$^{2}$, Suqi Cheng$^{2}$, Shuaiqiang Wang$^{2}$, Zhicong Cheng$^{2}$, Shuzi Niu$^{3*}$, and Dawei Yin$^{2*}$}
\affiliation{\institution{$^1$ University of Chinese Academy of Sciences, Beijing, China} \country{}}
\affiliation{ \institution{$^2$ Baidu Inc., Beijing, China} \country{}}
\affiliation{ \institution{$^3$ Institute of Software, Chinese Academy of Sciences, Beijing, China} \country{}}
\email{dongqian19@mails.ucas.ac.cn,chengzhicong01@baidu.com, shuzi@iscas.ac.cn, yindawei@acm.org}
\email{{liuyiding.tanh, chengsuqi, shqiang.wang}@gmail.com}

\thanks{$^{\dagger}$ This work wad done during Qian Dong's internship at Baidu.}
\thanks{$^{*}$ Co-corresponding authors.}

\begin{abstract}
Passage re-ranking is to obtain a permutation over the candidate passage set from retrieval stage. Re-rankers have been boomed by Pre-trained Language Models (PLMs) due to their overwhelming advantages in natural language understanding. However, existing PLM based re-rankers may easily suffer from vocabulary mismatch and lack of domain specific knowledge. To alleviate these problems, explicit knowledge contained in knowledge graph is carefully introduced in our work. Specifically, we employ the existing knowledge graph which is incomplete and noisy, and first apply it in passage re-ranking task. To leverage a reliable knowledge, we propose a novel knowledge graph distillation method and obtain a knowledge meta graph as the bridge between query and passage. To align both kinds of embedding in the latent space, we employ PLM as text encoder and graph neural network over knowledge meta graph as knowledge encoder. Besides, a novel knowledge injector is designed for the dynamic interaction between text and knowledge encoder. Experimental results demonstrate the effectiveness of our method especially in queries requiring in-depth domain knowledge.

\end{abstract}

\ccsdesc[500]{Information systems~Language models}
\ccsdesc[500]{Information systems~Learning to rank}
\ccsdesc[500]{Information systems~Similarity measures}
\ccsdesc[300]{Information systems~Novelty in information retrieval}

\keywords{Learning to Rank; Language models; Semantic Matching}

\maketitle

\section{Introduction}
\emph{Passage Re-ranking} is a crucial stage in modern information retrieval systems, which aims to reorder a small set of candidate passages to be presented to users. To put the most relevant passages on top of a ranking list, a re-ranker is usually designed with powerful capacity in modeling semantic relevance, which attracted a wealth of research studies in the past decade~\cite{guo2020deep}. Recently, 
large-scale pre-trained language models (PLMs), e.g. BERT~\cite{devlin2018bert}, ERNIE~\cite{sun2019ernie} and RoBERTa~\cite{liu2019roberta}, have dominated many natural language processing tasks, and have also achieved remarkable success on passage re-ranking.
For example, PLM based re-rankers~\cite{macavaney2019cedr,li2020parade,dong2021latent,dong2022disentangled} have achieved state-of-the-art performance, which takes the concatenation of query-passage pair as input, and applies multi-layer full-attention to model their semantic relevance. Their superiority can be attributed to the expressive transformer structure and the pretrain-then-finetune paradigm, which allow the model to learn useful implicit knowledge (i.e., semantic relevance in the latent space) from massive textual corpus~\cite{fan2021pre}. 

However, \emph{implicit knowledge} still has some inherent weaknesses, which limits the applicability of PLMs based re-rankers. First, 
queries and passages are usually created by different persons and have different expression ways~\cite{nogueira2019document}, such as word usage and language style. 
Worse still, the data distributions of search queries and web contents are highly heterogeneous~\cite{liu2021pre}, where various specialized domains (e.g., bio-medical) may only have few training examples in a general corpus. Domain-specific knowledge can hardly be revealed and captured by the model, and thus the processing of domain-specific queries is often inaccurate.

\begin{figure*}
		\centering
		\includegraphics[width=0.8\linewidth]{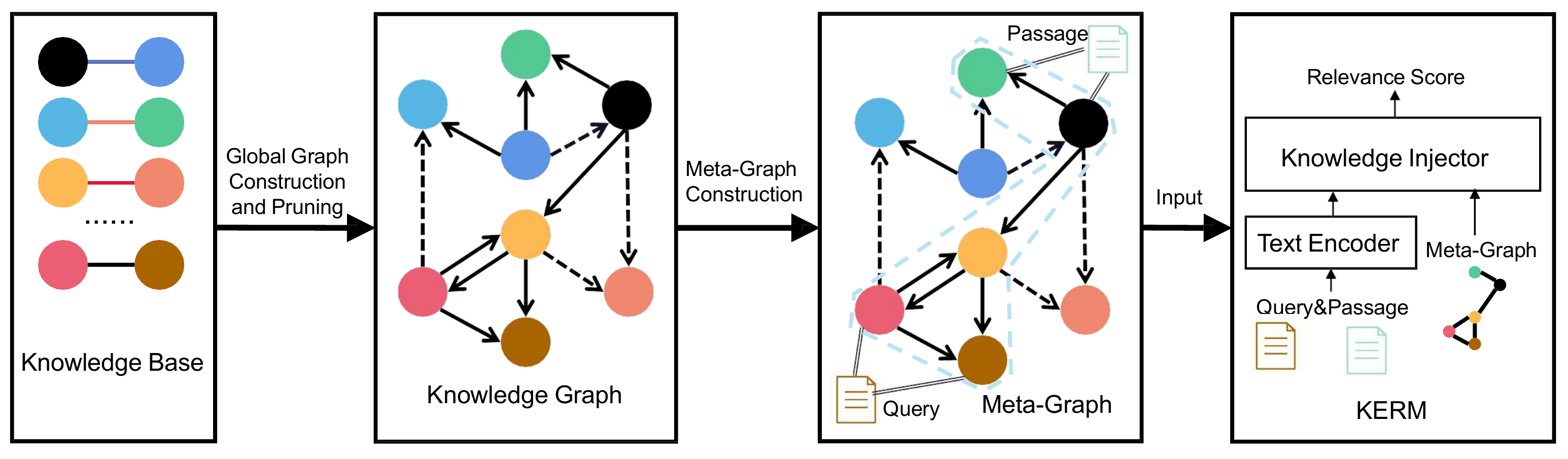}
		\caption{The workflow of KERM.}
		\label{fig:workflow}
\end{figure*}

To overcome the limitations, it is essential to incorporate the knowledge graph as \emph{explicit knowledge} to PLM based re-rankers. Thus we propose \textbf{K}nowledge \textbf{E}nhanced \textbf{R}e-ranking \textbf{M}odel (\textbf{KERM}), which utilizes external knowledge to \emph{explicitly} enhance the semantic matching process in PLM based re-rankers.
Intuitively, the difference in expression ways can be mitigated by the triplet with "synonymy" as relation in knowledge graph, and all the triplets can enrich the domain knowledge.
The overall workflow of KERM is depicted in Fig.~\ref{fig:workflow}.
To the the best of our knowledge, this is the first attempt for knowledge enhanced PLMs for passage re-ranking.


Despite the knowledge graph is a desirable source of explicit knowledge, it is non-trivial to take advantage of explicit knowledge directly for passage re-ranking due to the following two challenges:

\begin{itemize}[leftmargin=*]
\item \textbf{Challenge 1.} Existing knowledge graph are not constructed for re-ranking task. They usually contain trivial factual triples, which can hardly bring information gain. The inappropriate selection of external knowledge could even jeopardize the re-ranker performance. How to utilize existing knowledge graph to re-ranking task is remain a challenge.
 
\item \textbf{Challenge 2.} 
The explicit knowledge and implicit knowledge are highly heterogeneous due to the different sources, which makes the aggregation of the two difficult.
How to mutually refine each other and effectively aggregate explicit knowledge into implicit knowledge to alleviate the semantic gap between query and passage is still a challenge.
\end{itemize}


In general, the workflow of KERM can be divided into knowledge graph distillation and knowledge aggregation to tackle the above challenges.

For \emph{knowledge graph distillation}, we propose a novel pipeline to establish knowledge meta graphs, which only retain informative knowledge for passage re-ranking. Specifically, we first distill a graph globally for passage re-ranking scenario from an existing knowledge graph by pruning some unreliable or noisy relations based on TransE embedding. Then for a specific query-passage pair, we extract entities from both the query and passage, and construct a query-document bipartite entity graph based on query and passage entities and their k-hop neighbors, namely knowledge meta graph. \textbf{Challenge 1.} could be addressed in this distillation process.

For \emph{knowledge aggregation}, we design a novel interaction module between text and knowledge graph to combine the implicit and explicit knowledge. To derive implicit knowledge from text, we employ PLM as text encoder. To be aligned with implicit knowledge, knowledge meta graph is encoded with a multi-layer graph neural network (i.e. k-hop), namely Graph Meta Network (GMN). Each transformer layer outputs word representations. Each graph meta network layer outputs entity representations. Both word and entity representations are aggregated as the input of the following transformer and GMN layer, respectively in a novelly designed module, namely knowledge injector. Therefore through knowledge aggregation, implicit knowledge from text corpus and explicit knowledge from existing knowledge graph can mutually boost each other to achieve a better re-ranking performance, in which the issues in \textbf{Challenge 2.} could be mitigated.

Overall, our contributions can be summarized as follows:
\begin{itemize}
\item It is the first attempt to solve the knowledge enhanced PLMs problem for passage re-ranking. The key motivation lies in that bridging the semantic gap between the query and passage with the help of both kinds of knowledge.

\item We design a novel knowledge graph distillation method. It refines a reliable knowledge graph from the existing one globally and constructs a knowledge meta graph based on the refined graph locally.

\item We propose a novel aggregation of PLM and graph neural network framework to model the interaction between explicit knowledge and implicit knowledge.

\item Experimental results show the effectiveness of KERM on both general and domain specific data, achieving state-of-the-art performance for passage re-ranking. We also conduct a comprehensive study for the effects of each module in our method. The code is available at https://github.com/DQ0408 /KERM.
\end{itemize}

\section{Related Work}
\label{sec:relatedwork}
In this section, we introduce several recently-proposed PLMs based re-rankers and retrievers. Moreover, we also present the general background of the related techniques involved in this paper, i.e. Knowledge Enhanced Pre-trained Language Models (KE-PLMs) and Graph Neural Network.

\subsection{PLMs based Re-rankers}
\label{sec:plmreranker}
Existing PLMs based re-rankers typically improve ranking performance from two aspects: (1) \textbf{By optimizing the ranking procedure}: monoBERT~\cite{nogueira2019passage} is the first work that re-purposed BERT as a passage re-ranker and achieves state-of-the-art results. duoBERT~\cite{nogueira2019multi} integrates monoBERT in a multistage ranking architecture and adopts a pairwise classification approach to passage relevance computation. UED~\cite{yan2021unified} proposes a cascade pre-training manner that can jointly enhance the retrieval stage through passage expansion with a pre-trained query generator and thus elevate the re-ranking stage with a pre-trained transformer encoder. The two stages can facilitate each other in a unified pre-training framework. H-ERNIE~\cite{chu2022} proposes a multi-granularity PLM for web search.
(2) \textbf{By designing rational distillation procedure}: LM Distill + Fine-Tuning~\cite{gao2020understanding} explores a variety of distillation methods to equip a smaller re-ranker with both general-purpose language modeling knowledge learned in pre-training and $\textit{search-}$ $specific$ relevance modeling knowledge learned in fine-tuning, and produces a faster re-ranker with better ranking performance. CAKD~\cite{hofstatter2020improving} proposes a cross-architecture knowledge distillation procedure with a Margin-MSE loss, which can distill knowledge from multiple teachers at the same time. RocketQAv1~\cite{qu2021rocketqa} trains dual-encoder and cross-encoder in a cascade manner, which leverages the powerful cross-encoder to empower the dual-encoder. RocketQAv2~\cite{ren2021rocketqav2} proposes a novel approach that jointly trains the dense passage retriever and passage re-ranker. The parameters of RocketQAv2 are inherited from RocketQAv1. Besides, RocketQAv2 utilizes a large PLM for data augmentation and denoising, which can also be regarded as a distillation procedure. Notably, these two types of studies anticipate more insightful information to be captured by the advanced ranking and training procedures, while neglecting the limitations of implicit knowledge extracted from noisy and heterogeneous data. Therefore, in this paper, we proposed the first knowledge-enhanced PLM based re-ranker, which thoughtfully leverages explicit external knowledge that improve the effectiveness of the model.

\subsection{PLMs based Retrievers}
The low-dimensional dense representations for query and passage are computed by PLMs based retrievers from the dual-encoder architecture. Afterward, the candidate passage set could be retrieved efficiently via approximate nearest neighbor algorithms. 
Existing studies could be categorized into two parts: 
(1) By optimizing the matching stage: DPR~\cite{karpukhin2020dense} is the first study to leverage PLM to empower the retriever by a single vector. Other researches, such as
RepBERT~\cite{zhan2020repbert}, ColBERT~\cite{khattab2020colbert}, COIL~\cite{gao2021coil} and Interactor~\cite{ye2022fast}, obtain multiple vectors for query and passage for matching.
(2) By optimizing the representation learning module: RocketQAv1~\cite{qu2021rocketqa} and RocketQAv2~\cite{ren2021rocketqav2} boost the representation learning of retriever by leveraging the power of cross-encoder in a cascade or joint manner. Other studies boost the representation learning by designed IR-oriented pre-training tasks.
ICT~\cite{lee2019latent} treats sentences as pseudo-queries and matched them to the passage they originate from. Condenser~\cite{gao2021condenser} utilizes a novel pre-training task, which can produces an information-rich representation to condense an input sequence.
\subsection{Knowledge Enhanced Pre-trained Language Models (KE-PLMs)}
Existing KE-PLMs can be categorized by the granularity of knowledge they incorporate from knowledge graph (KG), as text-based knowledge, entity knowledge and KG meta-graphs.
To integrate text-based knowledge, RAG~\cite{lewis2020retrieval} and KIF~\cite{fan2020augmenting} first retrieve top-k documents from Wikipedia using KNN-based retrieval, and the PLM model is employed to generate the output conditioned on these retrieved documents. Entity-level information can be highly useful for a variety of natural language understanding tasks. Hence, many existing KE-PLMs target this type of simple yet powerful knowledge. ERNIE(BAIDU)~\cite{sun2019ernie} introduces a new pre-training strategy of language model which masking phrases or entities in order to implicitly learn both synaptic and semantic knowledge from these units. ERNIE(THU)~\cite{zhang2019ernie} integrates informative entity representations in the knowledge module into the underlying layers of the semantic module based on the alignments between text and entity to equip the model with the ability of knowledge awareness. As knowledge graphs provide richer information than simply entity, more and more researchers start to explore integration of more sophisticated knowledge, such as meta-graphs in KG. CokeBERT~\cite{su2021cokebert} proposes a novel semantic-driven Graph Neural Network (GNN) to dynamically select contextual knowledge and embed knowledge context according to textual context for PLMs, which can avoid the effect of redundant and ambiguous knowledge in KGs that cannot match the input text.
CoLake~\cite{sun2020colake} also uses GNN to aggregate information from the constructed meta-graph in both pre-training and inference. CoLake converts the meta-graph into token sequence and appends it to input sequence for PLMs, which is distinctive to CokeBERT. Although extensive research has been proposed up to now to address the knowledge-aware problem, none exists which constrained on how to use knowledge to empower PLMs particularly for re-ranking tasks.

\subsection{Graph Neural Network}
Existing Graph Neural Networks (GNNs) mainly fall into two categories: graph-based and path-based. Graph-based GNNs learn the structured information by directly passing nodes massage on the graph structure. 
GCNs~\cite{kipf2016semi} introduce a novel approach on graph-structured data by aggregating messages from its direct neighbors to learn the graph-structured feature efficiently and effectively. R-GCNs~\cite{schlichtkrull2018modeling} are developed specifically to encode the highly multi-relational graphs by defining relation-specific weight matrix for each edge type.
In contrast, path-based GNNs first decompose the graph into paths and then pass nodes massage on the path level, which can naturally utilize the relationship between neighbors to transmit messages. 
RNs~\cite{santoro2017simple} use MLPs to encode all paths in a graph and then pool the representation of paths to generate a global representation for the graph. KagNet~\cite{lin2019kagnet} is a combination of GCNs, LSTMs and a hierarchical path-based attention mechanism, which forms an architecture for modeling nondegenerate paths in a graph. In this work, we use path-based GNNs to formulate our GMN module for its good scalability on modeling relationship information in heterogeneous graphs.

\section{Problem Formulation}
\subsection{Passage Re-ranking}

Given a query $\textbf{q}$, passage re-ranking aims at ordering a set of $\varkappa$ passages, i.e., 
$\mathcal{P}=\left\{\textbf{p}_{\kappa}\right\}_{\kappa=1}^{\varkappa}$, which is usually retrieved from a large-scale passage collection by a retriever, e.g. BM25~\cite{yang2017anserini}, DPR~\cite{karpukhin2020dense} etc. In particular, a passage is a sequence of words $\textbf{p}=\{w_p\}_{p=1}^{|\textbf{p}|}$, where $|\textbf{p}|$ is the length of passage $\textbf{p}$. Similarly, a query is a sequence of words $\textbf{q}=\{w_q\}_{q=1}^{|\textbf{q}|}$. Note that a passage $\textbf{p}$ consists of $T$ sentences $\textbf{p}=\{\textbf{s}_\tau\}_{\tau=1}^{T}$.

Following a previous study~\cite{zou2021pre}, a desirable re-ranker is a scoring function $f^*(\cdot,\cdot)$ that maximizes the consistency between its predictions (denoted as $\hat Y_{\textbf{q}, \mathcal{P}} = \{f(\mathbf{q}, \mathbf{p}_\kappa)~|~\mathbf{p}_\kappa \in \mathcal{P}\}$) and the ground truth labels (denoted as $Y=\{y_\kappa\}_{\kappa=1}^{\varkappa}$), i.e.,

\begin{equation}
f^{*}=\max _{f} \mathbb{E}_{\{\textbf{q}, \mathcal{P}, Y\}}{\vartheta(Y, \hat Y_{\textbf{q}, \mathcal{P}})},
\label{eq:scorefunc}
\end{equation}
where $\vartheta$ is a ranking metric (e.g., MRR@10) that measures the consistency between the predictions and the labels.

\subsection{Explicit Knowledge Enhanced Passage Re-ranking}
A knowledge base is usually represented as a directed graph $\mathcal{G}= \{ \mathcal{E},\mathcal{R} \}$, where the node set $\mathcal{E}$ represents entities, and the edge set $\mathcal{R}$ is composed of relations between entities.
A triplet $(e_h,r,e_t)$ is the basic unit in the knowledge graph, where $e_h, e_t \in \mathcal{E}$ are head and tail entity respectively, and $r \in \mathcal{R}$ refers to their relations.
For example, $(apple, used\_{for}, eating)$ means that "apple is used for eating".

To leverage explicit knowledge in $\mathcal{G}$ for passage re-ranking, we anticipate building a novel knowledge-enhanced passage re-ranker, whose objective can be defined as
\begin{equation}
f^{*}=\max _{f} \mathbb{E}_{\{\textbf{q}, \mathcal{P}, Y\}}{\vartheta(Y, \hat Y_{\textbf{q}, \mathcal{P}, \mathcal{G}})},
\label{eq:scorefunc_kg}
\end{equation}
where $\hat Y_{\textbf{q}, \mathcal{P}, \mathcal{G}} = \{f(\mathbf{q}, \mathbf{p}_\kappa ~|~\mathcal{G})~|~\mathbf{p}_\kappa \in \mathcal{P}\}$, and $f(\mathbf{q}, \mathbf{p}_\kappa ~|~\mathcal{G})$ represents the ranking score that is aware of the explicit knowledge extracted from $\mathcal{G}$.

\section{KERM}

In this section, we introduce \textbf{K}nowledge \textbf{E}nhanced \textbf{R}e-ranking \textbf{M}odel (\textbf{KERM}), which leverages explicit knowledge that improves conventional cross-encoder for passage re-ranking.
Notably, the main challenges of incorporating explicit knowledge are to 1) distill a knowledge graph that is useful for re-ranking task, and 2) aggregate the explicit knowledge with the current implicit knowledge in an appropriate manner that can improve the overall performance. Hence our proposed approach is mainly composed of two parts, i.e., \textbf{knowledge graph distillation} and \textbf{knowledge aggregation}, to tackle two challenges respectively. In the rest of this section, we first describe how to distill a reliable knowledge graph globally and build a knowledge meta graph locally from it for a specific query-passage pair. Then, we present how to combine the distilled knowledge graph and existing text corpus to derive a knowledge enhanced passage re-ranker.



\subsection{Knowledge Graph Distillation}
Existing knowledge graphs are usually incomplete and noisy. It is unsuitable for direct introduction of them to the current model. Specially, there is no knowledge base particularly for passage re-ranking task. For example, \emph{ConceptNet}~\cite{speer2017conceptnet} is a general knowledge graph that contains common sense knowledge, where the information might not be useful for our passage re-ranking task. Therefore, it is critical for us to propose a knowledge graph distillation process from both global and local perspectives.
\begin{figure}
\centering
\includegraphics[width=\linewidth]{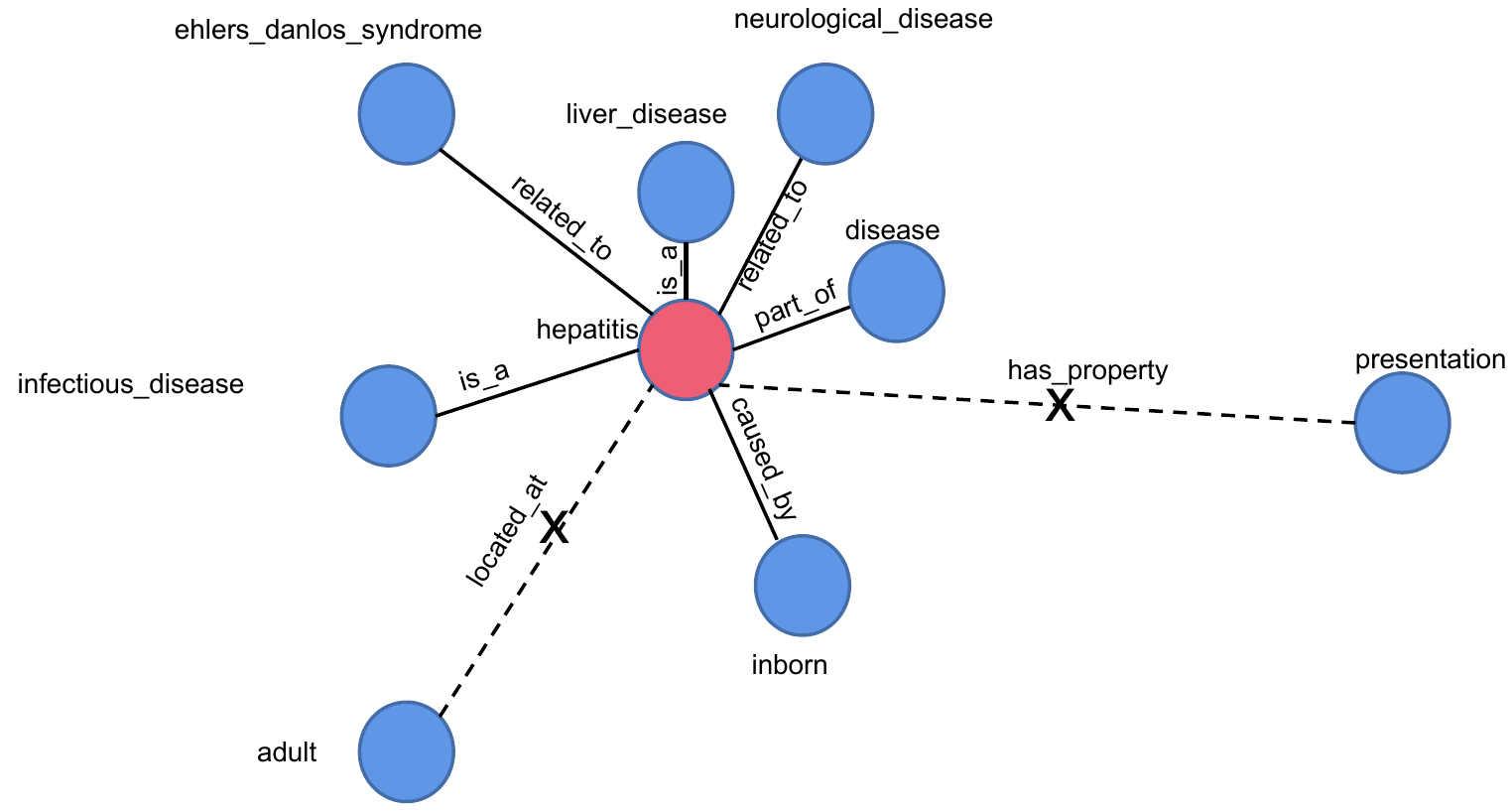}
\caption{The illustration of global graph pruning.}
\label{fig:Pruning}
\end{figure}
\subsubsection{Step 1: Global Graph Pruning}\hspace*{\fill} \\
Given a global knowledge graph $\mathcal{G}$, the first step is to eliminate those knowledge that might be noisy to be applied. To achieve this, we use TransE~\cite{bordes2013translating} to measure the reliability of a given knowledge triplet. In particular, TransE is an unsupervised learning method that learns latent representations for a knowledge triplet $(e_h,r,e_t)$. Intuitively, it models the latent distribution of knowledge in a given knowledge graph, and those who are out of this distribution can be viewed as less informative knowledge, which should not be used. Based on this, 
we use the entity embeddings pre-trained by TransE to calculate a distance metric between two linked entities as
\begin{equation}
Rel_{e}(e_h,r,e_t) = \mathbf{E}({e_h})\cdot \mathbf{E}(r)+\mathbf{E}({e_h})\cdot \mathbf{E}({e_t})+\mathbf{E}({r})\cdot \mathbf{E}({e_t}),
\label{eq:entrelfunc}
\end{equation}
\begin{equation}
Dist(e_h, e_t) = \frac{1}{Rel_{e}(e_h,r,e_t)},
\label{eq:entdistfunc}
\end{equation}
where $\mathbf{E}({e})$ and $\mathbf{E}({r})$ are the TransE embeddings of entity and relation, respectively, and the inner product measures the relevance between two vectors. As the objective of TranE is aligned with minimizing the distance shown in Eq.(\ref{eq:entdistfunc}), we can consider those knowledge triplets with small distance values as informative knowledge.

After measuring the reliability of knowledge, we prune $\mathcal{G}$ by only keep the top-$\Pi$ neighboring entities $\mathcal{N}(e_h)$ of a given entity $e_h$, which can formally be defined as
\begin{equation}
\mathcal{N}(e_h) = \cup_{\pi=1}^{\Pi}\{e_t^{\pi}\}, where\,Dist(e_h,e_t^{\pi})\leq Dist(e_h,e_t^{\pi+1}).
\label{eq:neighbor}
\end{equation}
Thus, the pruned global graph $\overline{\mathcal{G}}$ can be denoted as
\begin{equation}
\overline{\mathcal{G}} = \{(e_h,r,e_t)| e_h,e_t \in \mathcal{E}\wedge r \in \mathcal{R}\wedge e_t \in \mathcal{N}(e_h) \}.
\label{eq:prunedglobalgraph}
\end{equation}

Fig.~\ref{fig:Pruning} shows a real case of our global graph pruning method on ConceptNet, i.e., a general knowledge graph. In this case, the entity \emph{hepatitis} has various relations to \emph{disease}, \emph{infectious disease}, \emph{adult}, etc. From the distance of nodes in Fig.~\ref{fig:Pruning}, we can clearly observe that the knowledge \emph{hepatitis is an infectious disease} is more reliable and informative than \emph{hepatitis is located at adult}. To \emph{hepatitis}, the concept \emph{adult} is more general than \emph{infectious disease}. This indicates that our pruning method can effectively eliminate less informative knowledge.

\subsubsection{Step2: Meta-Graph Construction}\hspace*{\fill} \label{sec:meta-graph-construction}

\noindent Different from existing knowledge-enhanced PLMs for other NLP tasks, our aim for the re-ranking task is particularly on the relevance modeling between query and passage.
Thus, we further leverage the knowledge in the global graph $\overline{\mathcal{G}}$ to construct ``bridges'' between query and passage, which alleviates the semantic gap and improves semantic modeling.
More specifically, for a given query-passage pair (i.e., $(\mathbf{q}, \mathbf{p})$), we propose to construct a bipartite \emph{meta-graph} that connects those entities in the $\mathbf{q}$ and those in $\mathbf{p}$. 

\begin{algorithm}[!htp]
\caption{Meta-Graph Construction Algorithm}
\label{alg:construction}
\KwIn{query $\textbf{q}$, passage $\textbf{p}$ and pruned global graph $\overline{\mathcal{G}}$}
\KwOut{meta-graph $\textbf{G}$ of query $\textbf{q}$ and passage $\textbf{p}$}
\hrulefill 

\% key sentence selection;\\
$\textbf{s}^*=\arg\max_{\textbf{s}_i} Rel_{qs}(\textbf{q},\textbf{s}_i)$ as defined in Eq.(\ref{eq:qsrelfunc});\\
\% target entity recognition;\\
\FuncSty{${\mathrm{EntityRecognition}}(text)$} \Begin{
    $\phi=\{\}$;\\
    \For{$i\in[1..|text|]$}{
        \For{$length\in[max\_phrase\_length..1]$}{ 
            \If {$text[i:i+length] \in \mathcal{E}$} {
                $\phi = \phi \cup \{text[i:i+length]\}$;\\
                break;
            }
        }
    }
    return $\phi$;
}
$\phi_{\textbf{q}}=\mathrm{EntityRecognition}(\textbf{q})$,$\phi_{\textbf{s}^*}=\mathrm{EntityRecognition}(\textbf{s}^*)$;\\
\% meta-graph construction via path discovery;\\
k=1;$\textbf{G}=\{\}$;$\textbf{queue}=\{\}$;\\
\For{$e_h \in \phi_{\textbf{q}}$}{
    $\textbf{queue}=\textbf{queue}\cup \{(e_h,)\}$
}
\While{ $k\le K$}{
$\textbf{nextHop}=\{\}$;\\
\For{$path \in \textbf{queue}$}{
    $e_h=path[-1];$\\
    \For{$e_t \in \mathcal{N}(e_h)$}{
        \uIf{$e_t \in \phi_{\textbf{s}^*}$}{
            $\textbf{G} = \textbf{G} \cup \{path+(r,e_t)\}$;
        }
        \Else{
            $\textbf{nextHop} = \textbf{nextHop} \cup \{path+(r,e_t)\}$;
        }
    }
}
k+=1;$\textbf{queue}=\textbf{nextHop}$;\\
};\\
\end{algorithm}

The construction process is shown in Alg.~\ref{alg:construction}, which contains three sub-steps: key sentence selection, target entity recognition and path discovery.

\begin{enumerate}[leftmargin=*]
    \item \textbf{Key sentence selection}. The actual information need of a user usually concentrates on a small part of a relevant passage~\cite{guo2020deep}. To this end, we mimic human judgment and only focus on the sentence of each passage that is the most related to a query~\cite{zou2021pre}.
    In particular, we define the relevance score between a query $\textbf{q}$ and a sentence $\textbf{s}_i$ as 
    \begin{equation}
    Rel_{qs}(\textbf{q}, \textbf{s}_i) = \frac{\sum_{q=1}^{|\textbf{q}|}\textbf{E}(w_q)}{|\textbf{q}|} \cdot \frac{\sum_{s=1}^{|\textbf{s}_i|}\textbf{E}(w_s)}{|\textbf{s}_i|}.
    \label{eq:qsrelfunc}
    \end{equation}
    
    For the sake of efficiency, we initialize $\textbf{E}(w)$ from Word2Vec~\cite{mikolov2013linguistic} embedding.
    Based on Eq.(\ref{eq:qsrelfunc}), we select the most relevant sentence $\textbf{s}^*$ in $\textbf{p}$ to build the meta-graph for $\mathbf{q}$ and $\mathbf{p}$. 
    \item \textbf{Target entity recognition}. 
    Next, we select the entities in $\textbf{q}$ and $\textbf{s}^*$ to construct the meta-graph. Specifically, we only consider the entities that exactly match in $\mathcal{E}$. Meanwhile, we omit those entity phrases that are sub-sequences of other recognized entities. 
    For example, in the query "what causes low liver enzymes", both "liver" and "liver enzyme" are entities, but the entity "liver enzyme" is more informative to be recognized as the target entity, and "liver" should be omitted.
    \item \textbf{Path discovery}. Finally, given the target entities of $\textbf{q}$ and $\textbf{s}^*$ (denoted as $\phi_{\mathbf{q}}$ and $\phi_{\mathbf{s}^*}$, respectively), we perform Breadth First Search (BFS) on $\overline{\mathcal{G}}$ to discover the paths within $K$-hop between $\phi_{\mathbf{q}}$ and $\phi_{\mathbf{s}^*}$. Note that we only keep the within-$K$-hop paths that might be the most useful for the downstream re-ranking task. Meanwhile, the knowledge could be complemented from the $K$-hop paths. 
    
\end{enumerate}

\begin{figure}
		\centering
		\includegraphics[width=\linewidth]{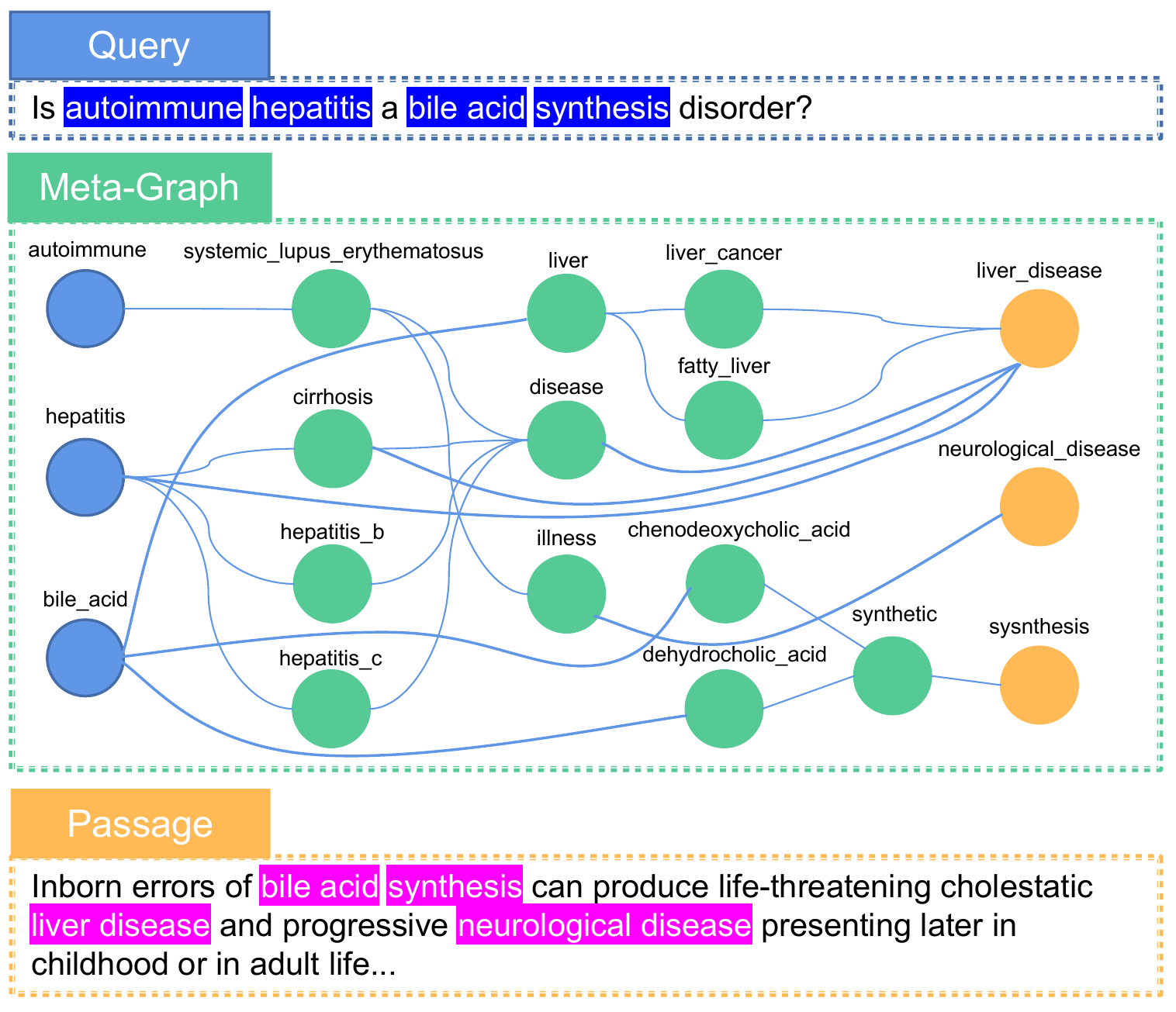}
		\caption{The illustration of a meta-graph constructed between the target entities of a query-passage pair. Note that some entities do not appear in the meta-graph, as they might not have any path that reaches the entities on the other side.}
		\label{fig:example}
\end{figure}

After taking the series of processes, the meta-graph $\mathbf{G}_{\mathbf{q}, \mathbf{p}}=\{\mathcal{E}_{\mathbf{q}, \mathbf{p}},\mathcal{R}_{\mathbf{q}, \mathbf{p}}\}$ is constructed with the multi-hop paths discovered between $\phi_{\mathbf{q}}$ and $\phi_{\mathbf{s}^*}$. Fig. \ref{fig:example} shows an example of the meta-graph, which contains rich knowledge about the semantic relevance between the query and passage. Notably, a better key sentence selector or entity linker such as Sentence-BERT~\cite{reimers2019sentence} and DER~\cite{wu2019zero} may benefit the ranking performance, but can burden the entire model inference time, which is infeasible to a qualified re-ranker. 

\begin{figure*}
		\centering
		\includegraphics[width=\textwidth]{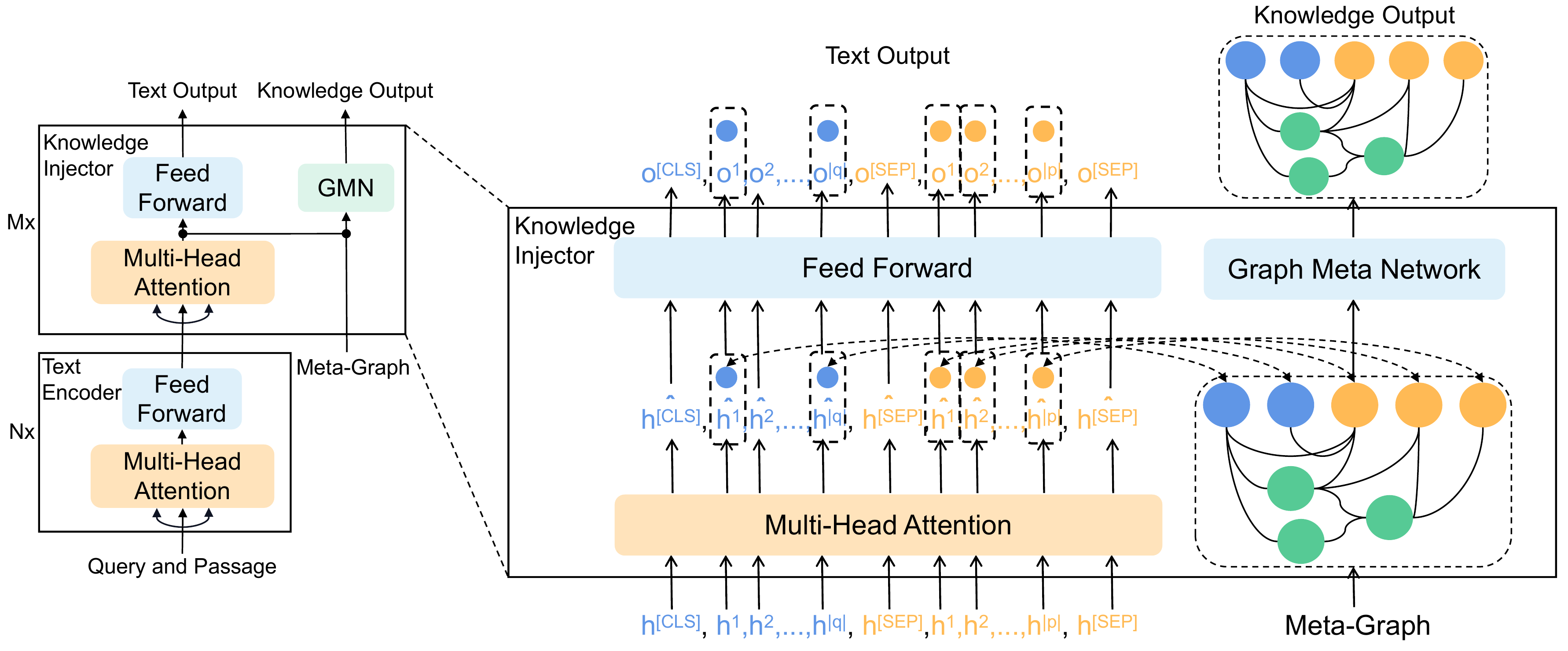}
		\caption{The architecture of KERM.}
		\label{fig:architecture}
\end{figure*}

\subsection{Knowledge Aggregation}
Given a meta-graph $\mathbf{G}_{\mathbf{q}, \mathbf{p}}$, we propose a PLM based re-ranker that performs knowledge-enhanced relevance computation, i.e., $f(\mathbf{q}, \mathbf{p} ~|~\mathcal{G})$. In the following, we first introduce the text encoder, and then present how we inject explicit knowledge from $\mathbf{G}_{\mathbf{q}, \mathbf{p}}$ into the encoder.

\subsubsection{Text Encoder}\hspace*{\fill} \\
We adopt the commonly-used cross-encoder as the text encoder. The input is formulated as 
the concatenation of a query-passage pair and the input layer converts the token indexes to a set of token embeddings~\cite{vaswani2017attention} (i.e., $\mathbf{O}_0$) as 
\begin{equation}
\mathbf{O}_0=\text{InputLayer}([[CLS],\{w_q\}_{q=1}^{|\mathbf{q}|},[SEP], \{w_p\}_{p=1}^{|\mathbf{p}|},[SEP]]).
\end{equation}
In the $l$-th transformer layer, text context features are extracted via multi-head self-attention and Feed Forward Network (FFN) as
\begin{equation}
\hat{\mathbf{H}}_l = \operatorname{MultiHeadAttention}(\mathbf{O}_{l-1}),
\label{eq:MultiHeadAttention}
\end{equation}
\begin{equation}
\mathbf{O}_l=\sigma\left(\hat{\mathbf{H}_l} \mathbf{W}_{l}^1+b_{l}^1\right) \mathbf{W}_{l}^2+b_{l}^2,
\label{eq:ffn}
\end{equation}
where $\mathbf{W}_l$ and $b_l$ are the parameters of FFN and $\sigma$ is an activation function, and $\mathbf{O}_l$ is the output of layer $l$.

\subsubsection{Knowledge Injector}\hspace*{\fill} \\
Based on the text encoder, we develop a knowledge injector that can seamlessly integrate explicit knowledge. Moreover, inspired by CokeBERT~\cite{su2021cokebert}, our knowledge injector is equipped with a GMN module to dynamically refine the knowledge context on the basis of text context features learned by text encoder, which further improves the flexibility and usability of the knowledge enhancement. Besides, our method allows the text context and knowledge context to interact and mutually boost each other.

\noindent\textbf{Knowledge injection}.
As shown in Fig.~\ref{fig:architecture}, the knowledge injector consists of multiple transformer layers, which is the same as the text encoder. Given a query-passage pair $(\mathbf{q}, \mathbf{p})$, we first find the entities in $\mathbf{G}_{\mathbf{q}, \mathbf{p}}$ that can be enhanced by external knowledge. For these entities, we define $\mathbf{E}
$ as the knowledge embeddings to be applied in the knowledge injection layers, where $\mathbf{E}$ is initialized by TransE embeddings extracted from the pruned global graph $\overline{\mathcal{G}}$. 

Next, we align each entity with the first token of the corresponding phrase in the selected key sentence~\cite{zhang2019ernie}, and define the knowledge injection process as
\begin{equation}
\hat{\mathbf{H}}_l = \operatorname{MultiHeadAttention}(\mathbf{O}_{l-1}),
\label{eq:MultiHeadAttention_again}
\end{equation}
\begin{equation}
\mathbf{F}_l=\sigma\left((\hat{\mathbf{H}_l} \mathbf{W}_{l}^1+b_{l}^1) \oplus \Lambda(\mathbf{E} \mathbf{W}_{l}^3+b_{l}^3)\right),
\label{eq:fuse}
\end{equation}
\begin{equation}
\mathbf{O}_l = \mathbf{F}_l\mathbf{W}^2_l+b^2_l.
\label{eq:textoutput}
\end{equation}
In Eq. (\ref{eq:fuse}), $\oplus$ means element-wise addition and $\Lambda(\cdot)$ represents the alignment function maps the entities to the corresponding positions of the tokens. By doing this, the external knowledge $\mathbf{E}$ is integrated in the output $\mathbf{O}_l$ of the knowledge injection layer. The final relevance score of this query-passage pair is defined as
\begin{equation}
    f(\mathbf{q}, \mathbf{p} ~|~\mathcal{G}) = \sigma\left(\mathbf{O}_\textrm{M}^{\textrm{[CLS]}} \mathbf{W}^4+b^4\right).
    \label{eq:relscore}
\end{equation}


\noindent\textbf{Knowledge propagation via meta-graph}.
It is worth noting that, the above-defined knowledge injection process only leverages knowledge embeddings learned by TransE on the global graph $\overline{\mathcal{G}}$. Particularly, it lacks considering the knowledge that bridges the semantics between query and passage. To this end, we introduce a Graph Meta Network (GMN) module that refines knowledge with the constructed meta-graph $\mathbf{G}_{\mathbf{q}, \mathbf{p}}$, The multi-hop paths of $\mathbf{G}_{\mathbf{q}, \mathbf{p}}$ allow the knowledge to be propagated between query and passage, which can enhance the relevance signal to be captured by the model, and thus alleviate the semantic gap.

More specifically, each knowledge injection layer has a multi-layer GMN (as shown in Fig. \ref{fig:architecture}) to propagate knowledge on $\mathbf{G}_{\mathbf{q}, \mathbf{p}}$.
First, the input of GMN is formulated with the fused feature $\mathbf{F}_l$ as
\begin{equation}
\hat{\mathbf{E}}_l^{(0)} = \Gamma( \mathbf{F}_l\mathbf{W}^5_l+b^5_l),
\label{eq:gmninput}
\end{equation}
where $\Gamma$ represents the slice operation that extracts the fused information of the target entities in $\mathbf{G}_{\mathbf{q}, \mathbf{p}}=\{\mathcal{E}_{\mathbf{q}, \mathbf{p}},\mathcal{R}_{\mathbf{q}, \mathbf{p}}\}$, and thus $\hat{\mathbf{E}}_l^{(0)}$ consists of fused entities representation $\hat{\mathbf{E}}^{(0)}_{e_1},\hat{\mathbf{E}}^{(0)}_{e_2},...,
\hat{\mathbf{E}}^{(0)}_{e_{\Psi}}$, i.e., $\Psi=|\mathcal{E}_{\mathbf{q}, \mathbf{p}}|$. 

Next, in the $k$-th layer of GMN, an entity embedding $e_h$ is updated via an attentive aggregation from its neighbors $\mathcal{N}(e_h)$ as 
\begin{equation}
\hat{\mathbf{E}}_{e_{h}}^{(k)}=\hat{\mathbf{E}}_{e_{h}}^{(k-1)}+\sum_{e_t \in \mathcal{N}(e_h)}\mathbf{a}_{ht}^{(k)}\hat{\mathbf{E}}_{e_{t}}^{(k-1)}.
\label{eq:aggr}
\end{equation}
Here, $\mathbf{a}_{ht}^{(k)}$ is the attention value, which can be defined as
\begin{equation}
\mathbf{a}_{ht}^{(k)}=\frac{exp(\mathbf{m}_{h t}^{(k)})}{\sum_{e_n \in \mathcal{N}(e_h)} exp(\mathbf{m}_{h n}^{(k)})},
\label{eq:attnscore}
\end{equation}
and the logits $\mathbf{m}_{h t}^{(k)}$ is computed as
\begin{equation}
\begin{aligned}
\mathbf{m}_{h t}^{(k)}=\sigma &\left(\alpha\left(\hat{\mathbf{E}}_{e_{h}}^{(k-1)} \| \hat{\mathbf{E}}_{e_{t}}^{(k-1)}\right)+\beta\left(\hat{\mathbf{E}}_{e_{h}}^{(k-1)} \| \hat{\mathbf{E}}_{r_{h t}}^{(k-1)}\right)\right.\\
&\left.+\gamma\left(\hat{\mathbf{E}}_{r_{h t}}^{(k-1)} \| \hat{\mathbf{E}}_{e_{t}}^{(k-1)}\right)\right).
\end{aligned}
\label{eq:sendfunc}
\end{equation}
In Eq. (\ref{eq:sendfunc}), the functions $\alpha(\cdot)$, $\beta(\cdot)$ and $\gamma(\cdot)$
are full-connected layers, and $\cdot\|\cdot$ represents concatenation operation. 
%
%

By applying a $K$-layer GMN in each layer of the knowledge injector, the output entity representation $\hat{\mathbf{E}}_{e_{h}}^{(K)}$ can ensemble knowledge from all the $K$-hop neighbors. As described in Section \ref{sec:meta-graph-construction} that all the paths of $\mathbf{G}_{\mathbf{q}, \mathbf{p}}$ between $\mathbf{q}$ and $\mathbf{p}$ is within $K$ hops, the GMN module can attentively propagate knowledge along the paths from entities in $\mathbf{p}$ to those in $\mathbf{q}$, and vice versa, which can enrich the semantics of the entities that benefit the relevance modeling.

Subsequently, the updated entity embeddings could be used as the knowledge to be injected in the next layer, i.e., $\mathbf{E} := \hat{\mathbf{E}}^{(K)}$. In other words, we can re-define Eq. (\ref{eq:fuse}) as
\begin{equation}
\mathbf{F}_l=\sigma\left((\hat{\mathbf{H}_l} \mathbf{W}_{l}^1+b_{l}^1) \oplus \Lambda(\mathbf{E}_l \mathbf{W}_{l}^3+b_{l}^3)\right),
\label{eq:fuse_new}
\end{equation}
where $\mathbf{E}_l$ is defined as
\begin{equation}
    \mathbf{E}_l =  
\begin{cases}
    \hat{\mathbf{E}}_{l-1}^{(K)}, & l \in [2, M] \\
    \text{TransE embeddings}.              & l=1
\end{cases}    
\end{equation}


\subsection{Model Optimization}
\noindent\textbf{Knowledge-enhanced pre-training}.
Following previous studies~\cite{nogueira2019multi,yan2021unified,kim2021self}, we conduct continual pre-training on MSMARCO corpus to warm up the parameters of GMN module.
We apply Masked Language Model (MLM)~\cite{devlin2018bert} and Sentence Relation Prediction (SRP)~\cite{wang2019structbert} as the pre-training tasks in KERM. 
Compared to conventional Next Sentence Prediction (NSP)~\cite{devlin2018bert}, the task of SRP is to predict whether a given sentence is the next sentence, previous sentence relation or no relation with another sentence. To incorporate knowledge during the pre-training stage, we construct a meta-graph for each sentence pair, and apply the knowledge aggregation process as introduced above.
The pre-training loss is defined as
$\mathcal{L}_{p}=\mathcal{L}_{MLM}+\mathcal{L}_{SRP}$.

\noindent\textbf{Knowledge-enhanced fine-tuning}.
We adopt a cross-entropy loss to fine-tune KERM:
\begin{equation}
\mathcal{L}_{f}=-\frac{1}{|\mathcal{Q}|} \sum_{q \in \mathcal{Q}} \log \frac{\mathrm{exp}({f(\mathbf{q}, \mathbf{p}^{+} ~|~\mathcal{G})})}{\mathrm{exp}({f(\mathbf{q}, \mathbf{p}^{+} ~|~\mathcal{G})})+\sum_{p^{-}} \mathrm{exp}({f(\mathbf{q}, \mathbf{p}^{-} ~|~\mathcal{G})})}
\end{equation}
where $|\mathcal{Q}|$ is the number of queries in training set, and $p^+$ and $p^{-}$ denote the positive passage and negative passage in $\mathbb{P}$ for current query $\mathbf{q}$, respectively.
\section{Experimental Setting}
%

\subsection{Datasets}
We use a large-scale public available corpus, i.e., MSMARCO-Passage collection~\cite{nguyen2016ms}, as our passage collection. This collection contains approximately 8.8 million passages extracted from 3.2 million web documents covering multiple fields. We train our model on the MSMARCO-TRAIN query set of 502,939 queries and evaluate KERM on three query sets. Table \ref{tab:datasets} provides the detailed information of these query sets.
The first test set is \textbf{MSMARCO-DEV}, which includes 6,980 sparsely-judged queries mixed with multiple domains. Each query has an average of 1.1 relevant passages with binary relevance label.
The second test set is \textbf{TREC 2019 DL}~\cite{craswell2020overview}, which contains 43 densely-judged queries with fine-grained relevance labels, i.e., irrelevant, relevant, highly relevant and perfectly relevant. On average, a query has 95.4 relevant passages, and most queries have more than 10 relevant passages. With fine-grained labels and multiple relevant passages per query, TREC 2019 DL can be used to reflect the fine-grained ranking performance between relevant passages. 
To evaluate KERM on specific domains, we further introduce \textbf{Ohsumed}~\footnote{http://disi.unitn.it/moschitti/corpora.htm} query set, which contains 63 queries on bio-medical domain. 
The collection of Ohsumed is constructed from the first 20,000 passages in Mesh categories of the year 1991. 
Following the previous work~\cite{joachims1998text}, the test collection including 10,000 passages are utilized for performance comparison on Ohsumed query set.
Each query has an average of 50.9 relevant passages with three graded relevance labels. In section~\ref{subsec:topics}, we demonstrate that the quality of external knowledge constructed by KERM in such domain could be more useful.

We use \textbf{ConceptNet}~\cite{speer2017conceptnet}, a general knowledge graph as our external knowledge base $\mathcal{G}$. Following KagNet~\cite{lin2019kagnet}, we merge relation types to increase graph density and construct a multi-relational graph with 17 relation types, including $atlocation$, $causes$, $createdby$, etc.

\begin{table*}[!htp]
	\centering
	\caption{Statistics of datasets.}
	\label{tab:datasets}
	\begin{tabular}{l|c|c|c|c|c}
		\hline
		              &  \#Queries  & \#Rel.Psgs. & Rel.Psgs./Query &\#Graded.Labels & Domain \\
		\hline
		MSMARCO-DEV   & 6980        & 7437      &1.1              &2               & General\\
		\hline
		TREC 2019 DL  & 43          & 4102      &95.4             &4               & General\\
		\hline
		OHSUMED       & 63          & 3205      &50.9             &3               & Bio-Medical\\
		\hline
	\end{tabular}
\end{table*}



\subsection{Baselines}
We include several PLMs based re-rankers in our evaluation, including the state-of-the-art:
\begin{itemize}
    \item \textbf{monoBERT}~\cite{nogueira2019passage}: The first study that re-purposes BERT as a re-ranker and achieves state-of-the-art results.
    \item \textbf{duoBERT}~\cite{nogueira2019multi}:
    This work proposes a pairwise classification approach using BERT, which obtains the ability to be more sensitive to semantics through greater computation.
    \item \textbf{UED}~\cite{yan2021unified}: A unified pre-training framework that jointly refines re-ranker and query generator. For a fair comparison, we only use the re-ranker in UED without passage expansion.
    \item \textbf{LM Distill+Fine-Tuning (LDFT)}~\cite{gao2020understanding}: 
    A variety of distillation methods are compared in this paper. The experimental results indicate that a proper distillation procedure (i.e. first distill the language model, and then fine-tune on the ranking task) could produce a faster re-ranker with better ranking performance.
    \item \textbf{CAKD}~\cite{hofstatter2020improving}: This work proposes a cross-architecture knowledge distillation procedure with Margin-MSE loss, which can distill knowledge from multiple teachers.
    \item \textbf{RocketQAv1}~\cite{qu2021rocketqa}: This work mainly focuses on the training of PLM based retriever, where the re-ranker is an intermediate product of its training process.
    \item \textbf{RocketQAv2}~\cite{ren2021rocketqav2}: Based on RocketQAv1, this work proposes a novel approach that jointly trains the PLM based retriever and re-ranker.
\end{itemize}
To compare the performance of different methods, we resort to two ranking metrics.
For MSMARCO-DEV, We adopt Mean Reciprocal Rank (i.e., MRR@10). 
For TREC 2019 DL, we use Mean Average Precision, i.e., MAP@10 and MAP@30.
For Ohsumed, both Mean Reciprocal Rank and Mean Average Precision (i.e., MRR@10 and MAP@10) are employed for comprehensive performance analysis in queries requiring in-depth domain knowledge.

\subsection{Implementation Details}
We use the traditional sparse retriever BM25~\cite{yang2017anserini} as our first stage method. All experiments are conducted under the \textbf{same BM25 setting} with 1000 retrieved candidates. We conduct experiments with the deep learning framework PaddlePaddle~\cite{ma2019paddlepaddle} on up to 4 NVIDIA Tesla A100 GPUs (with 40G RAM). For the GMN module, we use Paddle Graph Learning (PGL)~\footnote{https://github.com/PaddlePaddle/PGL}, an efficient and flexible graph learning framework based on PaddlePaddle. For training, we used the Adam optimizer~\cite{kingma2014adam} with a learning rate of 1e-5 for text encoder and 1e-4 for knowledge injector. The model is trained up to 5 epochs with a batch size of 640 and 240 for base and large models respectively. 
In our experiments, the PLM small, base and large models have 6, 12 and 24 Transformer layers respectively.
The text encoder has 9 layers and 21 layers for base and large model respectively, and the knowledge injector both has 3 layers in our experiment. The dropout rates are set to 0.1. The ratio of the positive to the hard negative is set to 1:19. 
All transformer layers in KERM's backbone are initialized from ERNIE-2.0 base~\cite{sun2020ernie}, which is a BERT-like model pre-trained with a continual pre-training framework on multiple tasks. We perform Knowledge-enhanced pre-training on MARCO passage collection to warm up the parameters in knowledge injector, which has 60,000 iterations under the batch size of 256.
For a fair comparison, the same pre-training without knowledge enhancement is also conducted on $\textrm{ERNIE}_{\textrm{base}}$ re-ranker and all models in ablation studies.

\section{Experimental Results}
Here we compare ranking performances of KERM and other PLMs based re-rankers on the first two widely used query sets. Moreover, ablation studies for each component of KERM are also explored. All experimental results were reported under the same BM25 setting. 
\subsection{Effectiveness Analysis}
\begin{table*}[!htp]
\centering
\caption{Performance comparison on MARCO-DEV and TREC 2019 DL.}
\label{tab:performance}
\begin{threeparttable}
\begin{tabular}{l|c|c|c|c|cc}
\hline 
&\multicolumn{3}{c|} { Model Settings }& \multicolumn{1}{c|} { MARCO DEV Queries } & \multicolumn{2}{c} { TREC 2019 DL Queries }\\
\cline{2-7}
                              &PLM             &Distillation&$\textrm{Batch Size} \times \textrm{GPU}$&MRR@10& MAP@10 & MAP@30\\
\hline 
BM25(anserini)        & w/o              &     w/o & - &18.7     & 11.3 & 20.1\\
\hline 
monoBERT              &$\textrm{BERT}_{\textrm{large}}$  &    w/o  &  $\textrm{128}\times\textrm{1}$    &37.2& 16.5 &30.1\\
duoBERT\tnote{1}               &$\textrm{BERT}_{\textrm{large}}$ &    w/o  &   $\textrm{128}\times\textrm{1}$  &39.0& 16.8 & 30.9   \\
UED\tnote{2}&$\textrm{BERT}_{\textrm{large}}$ &    w/o   &  $\textrm{64}\times\textrm{4}$  &39.5& - & -   \\
\hline
LDFT&$\textrm{BERT}_{\textrm{small}}$&$\textrm{BERT}_{\textrm{base}}$&$\textrm{64}\times\textrm{4}$&35.6 & - & - \\
CAKD\tnote{3}                &$\textrm{BERT}_{\textrm{small}}$&Multiple $\textrm{BERT}_{\textrm{large}}$&$\textrm{32}\times\textrm{1}$&39.1& 15.8 &30.6\\
RocketQAv1         &$\textrm{ERNIE}_{\textrm{large}}$  &w/o& $\textrm{64}\times\textrm{4}$ & 39.2 & 16.6  & 30.9  \\
RocketQAv2      &$\textrm{ERNIE}_{\textrm{base}}$  &$\textrm{ERNIE}_{\textrm{large}}$  &$\textrm{384}\times\textrm{32}$&\textbf{40.1}&   16.5 & 31.8   \\
\hline
$\textrm{ERNIE}_{\textrm{base}}$                &$\textrm{ERNIE}_{\textrm{base}}$  &    w/o     &  $\textrm{160}\times\textrm{4}$  &38.9&    15.6 & 29.5\\
$\textrm{KERM}_{\textrm{base}}$                          &$\textrm{ERNIE}_{\textrm{base}}$+GMN&  w/o     &  $\textrm{160}\times\textrm{4}$  &39.6&   {17.2}&31.4\\
$\textbf{KERM}_{\textrm{large}}$                          &$\textrm{ERNIE}_{\textrm{large}}$+GMN&  w/o    &  $\textrm{60}\times\textrm{4}$   &\textbf{40.1} &  \textbf{17.8}  &\textbf{33.2}\\
\hline
\end{tabular}
\begin{tablenotes}
       \footnotesize
       \item[1] duoBERT has one more re-ranking stage than the others. 
       \item[2] Only UED's re-ranker is used for report without passage expansion.
       \item[3] CAKD uses a slightly better BM25 retriever on MARCO-DEV in its experiments.
     \end{tablenotes}
   \end{threeparttable}
\end{table*}

Table~\ref{tab:performance} shows the ranking performance of KERM and baselines on MSMARCO-DEV and TREC 2019 DL. In the second column, model settings are displayed, including the PLMs used in models, whether distillation is enabled and computing resources required for model training. From Table~\ref{tab:performance}, we observe the following phenomena.

(1) Compared with the best SOTA methods on the sparsely-judged MARCO-DEV query set, KERM outperforms all other baseline models except RocketQAv2.
It utilizes a well-trained cross-encoder $\textrm{ERNIE}_{\textrm{large}}$ in RocketQAv1 to remove the predicted negatives with low confidence scores and include the predicted positives with high confidence scores. This can be regarded as a distillation. Meanwhile, RocketQAv2 achieves promising performance through a very large batch size on enormous computational resources, which is hardly comparable to our technique that only requires up to 4 GPUs. In addition to RocketQAv2, both $\textrm{KERM}_{\textrm{base}}$ and $\textrm{KERM}_{\textrm{large}}$ exceed strong baseline models, including duoBERT with sophisticated multiple re-ranking stages and CAKD distilled from multiple large models. It demonstrates the effectiveness of external knowledge injection.

(2) Among both kinds of baselines, $\textrm{KERM}_{\textrm{large}}$ achieves the best performance on the densely-judged TREC 2019 DL query set. MAP @10 and MAP@30 measure the quality of the ranking result over related passages. Baseline models with larger networks usually perform better in MAP, which indicates that complex structure helps model capture the fine-grained differences between related passages. With the well-designed GMN module and introduced reliable external knowledge, $\textrm{KERM}_{\textrm{base}}$ achieves the best performance on MAP@10 even compared to various large baseline models.

(3) Distilled models typically perform better at putting the relevant passage at top positions, but the subtle differences between relevant passages cannot be captured effectively through relatively small distilled models. On the MARCO-DEV query set, LDFT~\cite{gao2020understanding} performs better than duoBERT on MRR@10 and the former model size is much smaller than the latter. It shows that distillation plays a great role in performance improvement. Because LDFT~\cite{gao2020understanding} neither release the code nor report MAP in the original paper, we omit its result on TREC 2019 DL query set. Additionally, models that perform well on MAP do not lead in MRR and vice versa, demonstrating that two metrics are to measure different aspects of the ranking quality. KERM shows the most stable performance on both metrics among all baseline models.

(4) Compared with $\textrm{ERNIE}_{\textrm{base}}$ we trained, $\textrm{KERM}_{\textrm{base}}$ shows a significant improvement on both two query sets. This indicates the explicit introduction of external knowledge can alleviate the semantic gap and heterogeneity between query and passage, and improve the semantic matching performance.

\subsection{Ablation Study for Knowledge Injector}
\label{subsec:injector}

Knowledge injector module including knowledge injection and propagation process realized as Graph Meta Network (GMN), is mainly responsible for the interaction between text and knowledge graph. To explore their roles in the ranking performance, we remove the knowledge injection, aggregation process and the whole module separately and keep other units unchanged in KERM. Experimental results of three base models are shown in Table~\ref{tab:injector}. KERM without knowledge injector module is degraded to vanilla $\textrm{ERNIE}$. KERM without knowledge propagation process is formally equivalent to ERNIE(THU)~\cite{zhang2019ernie}. KERM without knowledge injection process takes the text of query-passage pair and meta graph as separate inputs, and then concatenate two parts of outputs to feed into a linear layer by redefining Eq.(\ref{eq:fuse_new}) and Eq.(\ref{eq:relscore}) respectively as
\begin{equation}
    \mathbf{F}_l =  
\begin{cases}
    \sigma\left(\hat{\mathbf{H}_l} \mathbf{W}_{l}^1+b_{l}^1\right), & \textrm{for}\;\textrm{Eq.}(\ref{eq:textoutput}) \\
    \sigma\left(\mathbf{E}_l \mathbf{W}_{l}^3+b_{l}^3\right),              & \textrm{for}\;\textrm{Eq.}(\ref{eq:gmninput})
\end{cases}    
\label{eq:withoutinteract}
\end{equation}
\begin{equation}
    f(\mathbf{q}, \mathbf{p} ~|~\mathcal{G}) = \sigma\left(\left(\mathbf{O}_\textrm{M}^{\textrm{[CLS]}}\| \mathbf{E}_{\textrm{M}}^{(K)} \right) \mathbf{W}^6+b^6\right).
    \label{eq:newrelscore}
\end{equation}

\begin{table}[!htp]
	\caption{Performance Comparisons between different settings of knowledge injector on MSMARCO DEV.}
	\label{tab:injector}
    \begin{small}
	\begin{tabular}{l|c}
		\hline
		               & MRR@10 \\
		\hline
		\textbf{KERM}  &  39.6          \\
		\hline
		\quad w/o Knowledge Interaction & 38.5 \\
		\hline
		\quad w/o Knowledge Propagation & 38.9 \\
		\hline
		vanilla ERNIE & 38.9 \\
		\hline
	\end{tabular}
    \end{small}
\end{table}

Table~\ref{tab:injector} shows the performance comparisons between different settings of knowledge injector, which is statistically significant. From this table, we can observe the following phenomena. (1) MRR@10 of KERM without interaction and propagation process decreases at least $1\%$ respectively. This indicates both knowledge interaction and propagation processes play an indispensable role in ranking performance. (2) The performance of KERM without propagation is comparable to vanilla ERNIE. Not only query and passage entities, but also their multi-hop neighbors are essential for the ranking performance. (3) MRR@10 of KERM without knowledge interaction drops the most. It suggests the simple and straightforward way to aggregate knowledge graph with text does not work in the passage re-ranking scenario. The text and knowledge graph need to be refined with each other mutually in the interaction, which will be further analyzed in detail as follows.

To further explore the text-knowledge interaction influence on the ranking performance, we compare ranking performances from KERM with different numbers of knowledge injector layers. All experiments in Table~\ref{tab:layers} are conducted with the same experimental settings except the number of knowledge injector layers (denoted as $M$). Note that in our setting, the number of text encoder layers $N$ plus $M$ is always $12$, i.e. the number of layers in $\textrm{ERNIE}_{\textrm{base}}$. No knowledge injector layer ($M=0$) represents the vanilla $\textrm{ERNIE}_{\textrm{base}}$ re-ranker without explicit knowledge enhancement.
\begin{table}[]
\caption{MRR@10 results with different number of layers in knowledge injector on MSMARCO.}
\label{tab:layers}
\begin{tabular}{l|llllll}
\hline
\multirow{2}{*}{}           & \multicolumn{6}{c}{\#Layers}                                  \\ \cline{2-7} 
                            & \multicolumn{1}{l|}{0}    & \multicolumn{1}{l|}{1}    & \multicolumn{1}{l|}{2}    & \multicolumn{1}{l|}{3}    & \multicolumn{1}{l|}{4}    & 5    \\ \hline
\multicolumn{1}{c|}{MRR@10} & \multicolumn{1}{l|}{38.9} & \multicolumn{1}{l|}{39.1} & \multicolumn{1}{l|}{39.4} & \multicolumn{1}{l|}{\textbf{39.6}} & \multicolumn{1}{l|}{39.4} & 39.2 \\ \hline
\end{tabular}
\end{table}

With the increase of $M$ in Table~\ref{tab:layers}, the ranking performance is not improved linearly. Instead, the performance achieves the best at $M=3$ and then falls down (statistically significant). This performance variation trend is contradictory to our intuition that the more injector layers, the deeper interaction between text and knowledge, and the more performance improvement is expected. The possible reason lies in that the knowledge injector layer makes pretrained parameters from $\textrm{ERNIE}_{\textrm{base}}$ not reusable, which means the implicit knowledge learned from large-scale is not applicable to these layers. Hence the number choice of the knowledge injector layer is somehow determined by the trade-off between implicit and explicit knowledge.

\subsection{Ablation Study for Knowledge Graph Distillation}
Knowledge graph distillation is performed in both global and local perspectives. To explore their roles in the ranking performance, we remove the graph pruning globally and sentence selection locally respectively, keep other settings unchanged, and derive KERM without graph pruning and sentence selection respectively. From results on TREC 2019 DL in Table~\ref{tab:measure3}, observations are listed as below. (1) Without global graph pruning, MRR@10 and the average edge score, calculated through Eq.(\ref{eq:entrelfunc}), decrease the most, and the time efficiency drops slightly. This indicates the original knowledge graph exists noise data that affect performance. (2) Without sentence selection, the time efficiency drops the most and the average edge score decreases slightly, which proves that not every sentence in a passage has a positive effect on semantic matching. Overall, knowledge graph distillation is significant to KERM.
\begin{table}[!htp]
	\caption{Comparisons on different settings of knowledge graph distillation on TREC 2019 DL.}
	\label{tab:measure3}
    \begin{small}
	\begin{tabular}{l|c|c|c}
		\hline
		               &Avg. Edge Score& MRR@10 & Time\\
		\hline
		\textbf{KERM}  &     54.6        & \textbf{98.4}&21.4ms \\
		\hline
		\quad w/o Graph Pruning &   51.4     &  96.1      &27.0ms\\
		\hline
		\quad w/o Sentence Selection & 53.1 &  97.2      &60.1ms\\
		\hline
	\end{tabular}
    \end{small}
\end{table}

\subsection{Quantitative Analysis on Different Domains}
\label{subsec:topics}
We further investigate the ranking effect of KERM on a specific domain. Specifically, we conduct experiments on OHSUMED from bio-medical field, and a bio-medical query subset of MSMARCO-DEV including $1,110$ queries. This query subset is derived from the mixed domain query set of MSMARCO-DEV by k-means clustering method~\cite{hartigan1979algorithm}, while the remaining subset with $5,870$ queires is denoted as the general domain subset. Performance comparisons between KERM and BM25, ERNIE are shown in Table~\ref{tab:performance2}.

Results are obtained from Table~\ref{tab:performance2}. (1) Poor ranking performances of all models on bio-medical domain indicates that it is more challenging in the data scarcity scenario, where textual data is not covered widely in the PLMs' pretraining datasets. (2) Compared with ERNIE, KERM has a higher relative improvement in bio-medical domain than general domain. This demonstrates that the incorporation of knowledge graph is more useful for a data scarcity domain. To verify this idea, we compare the size of knowledge meta graph used for different domains as follows.
%
\begin{table}[!htp]
	\centering
	\caption{Ranking performance comparison on different domains.}
	\label{tab:performance2}
\begin{tabular}{l|c|c|cc}
\hline 
& \multicolumn{2}{c|} { MARCO Dev Queries }& \multicolumn{2}{c} {\multirow{2}{*}{Ohsumed Queries}}\\
\cline{2-3}
                              & General & Bio-Medical &       &\\
\cline{2-5}
                              &MRR@10   &MRR@10       &MRR@10 &MAP@10\\
\hline 
BM25                          & 19.2   &    16.4    & 69.6  &9.5   \\
ERNIE                         & 38.5   &    30.5    & 79.7 &10.1   \\
KERM                          & 39.7   &    32.1     & 81.2 &11.0   \\
\hline
\end{tabular}
\end{table}

We quantify the knowledge desirability as the size of average knowledge meta graph used in one domain. Specifically, we use the average number of edges as the size and average edge score calculated through Eq.(\ref{eq:entrelfunc}) as the reliability of the knowledge meta graph. From Table \ref{tab:measure2}, we can see that the meta-graph constructed on Bio-Medical domains is better in terms of quantity and quality. It indicates that the external knowledge found on professional domains contains more information.

\begin{table}[!htp]
	\centering
	\caption{Quantitative analysis of knowledge in meta-graph on different domains.}
	\label{tab:measure2}
    \begin{small}
	\begin{tabular}{l|c|c}
		\hline
		&    Avg. Edge NO. & Avg. Edge Score\\
		\hline
		MARCO General  & 47.9& 50.9 \\
		\hline
		MARCO Bio-Medical  & 57.7 & 53.9 \\
		\hline
		Ohsumed  & 58.6& 53.5 \\
		\hline
	\end{tabular}
    \end{small}
\end{table}

\section{Conclusion}
The main goal of this paper is to reasonably introduce external knowledge graph to PLMs for passage re-ranking. We first design a novel knowledge meta graph construction method to distill reliable and query related knowledge from a general and noisy knowledge graph. The knowledge meta graph bridges the semantic gap between each query and passage. Then we propose a knowledge injector layer for mutually updating text and knowledge representations, which transformers word to entity representations for graph meta network, vice versa. Knowledge Enhanced Ranking Model is pretrained with Masked Language Model (MLM) Sentence Relation Prediction (SRP) [38] tasks, and fine-tuned with cross entropy loss function for passage re-ranking task. Experimental results on public benchmark datasets show the effectiveness of the proposed method compared with state-of-the-art baselines without external knowledge due to its first attempt. The role of each module in KERM is also comprehensively analyzed. Since this work was limited to the one-to-one meta-graph of a query-passage pair built online, continued efforts are needed to make knowledge enhancement more efficient for both retrieval and re-ranking stage.

\section{Future Works}
Despite that the knowledge graph distillation in our method is empirically shown to be effective for the final performance, the implementation of graph pruning and meta-graph construction is still based on simple heuristics. A more promising way of formulating a useful meta-graph is to jointly learn a graph generator with the reranker in an end-to-end fashion, which enables more flexibility.
Besides, it is currently infeasible to exploit the external knowledge in the retrieval stage, which needs to exhaustively build massive meta-graphs for a large scale of candidates. A further study could focus on how to use external knowledge in PLM based retriever. 

\begin{acks}
This research work was funded by the National Natural Science Foundation of China under Grant No.62072447.
\end{acks}
\bibliographystyle{ACM-Reference-Format}
\balance
\bibliography{sample-base}

\end{document}